\documentclass[cup6b]{cupbook}
\usepackage{graphicx}
\usepackage{natbib}
\usepackage{amsmath}
\title[Rome, Italy, 27--30 April 2009]
      {The coming of age of X-ray polarimetry}
\author{}
\date{}
\begin{document}
\pagenumbering{arabic}


\author[Schnittman \& Krolik]{Jeremy D.\ Schnittman (Johns Hopkins
  University) \and Julian H.\ Krolik (Johns Hopkins University)}
\chapter{X-ray Polarization from Black Holes in the Thermal State}

\abstract{We present new calculations of X-ray polarization from black hole
accretion disks in the thermally-dominated state, using a Monte-Carlo
ray-tracing code in full general
relativity. In contrast to many previously published studies, our
approach allows us to include returning radiation that is deflected by
the strong-field gravity of the BH and scatters off of the disk before
reaching a distant
observer.  Although carrying a relatively small fraction of the total
observed flux, the scattered radiation tends to be highly polarized
and in a direction perpendicular to the direct radiation.
We show how these new features of the polarization 
spectra may be developed into a powerful tool for
measuring black hole spin and probing the gas flow in the innermost
disk.}

\section{Introduction}

A recent flurry of new mission proposals has renewed interest in
X-ray polarization from a variety of astrophysical sources, hopefully
marking the ``Coming of Age of X-ray Polarimetry'' in the very near
future. The
Gravity and Extreme Magnetism SMEX ({\it GEMS}) mission\footnote{\tt
  heasarc.gsfc.nasa.gov/docs/gems}, for example, should be able to
detect a degree of polarization $\delta < 1\%$ for a flux of a few
mCrab (e.g.\ \cite{swank:08} and these proceedings). A similar
detector for
the International X-ray Observatory ({\it IXO}) could achieve
sensitivity roughly $10\times$ greater ($\delta < 0.1\%$;
\cite{jahoda:07, costa:08}, Alessandro Brez 
in these proceedings). In this talk, based on our recent paper
\cite{schnittman:09}, we focus on the polarization signal from
accreting stellar-mass black holes (BHs) in the thermal state, which are
characterized by a broad-band spectrum peaking around 1 keV. The
typical level of polarization from these sources should be a few
percent in the $1-10$ keV range, depending on BH spin and the
inclination angle of the accretion disk. 

Symmetry arguments demand that in the flat-space (Newtonian) limit, the
observed polarization from the disk must 
be either parallel or perpendicular to the BH/disk rotation axis.
However, the effects of relativistic beaming,
gravitational lensing, and gravito-magnetic frame-dragging can combine to give
a non-trivial net rotation to the integrated polarization
vector.  Early work exploring these effects
\cite{stark:77,connors:77,connors:80}
showed that they create changes in the angle and degree of polarization
that are strongest for higher photon energy. Quite recently, Dovciak
et al.\ \cite{dovciak:08}
investigated the effect of atmospheric optical depth on the
polarization signal, and Li et al.\ \cite{li:08} applied the original
calculations of thermal X-ray polarization to the problem of measuring
the inclination of the inner accretion disk. 

Nearly all previous work has modeled the relativistic effects by calculating
the transfer function along geodesics between the observer and
emitter. By its very nature, this method precludes the possibility of
including the effects of returning radiation---photons emitted from one
part of the disk and bent by gravitational lensing so that they are
absorbed or reflected from another part of the disk \cite{cunningham:76}.
As described in greater detail in \cite{schnittman:09}, the most important
feature of our approach is that the photons are traced {\it from} the
emitting region in all directions, either returning to the disk,
scattering through a corona, getting captured by the BH, or eventually
reaching a distant observer. 

Using this method, we study an important new polarization feature,
namely the transition between
horizontal- and vertical-oriented polarization as the photon energy
increases, an effect first discussed in \cite{agol:00}.
At low energies we reproduce the ``Newtonian'' result of
a semi-infinite scattering atmosphere emitting radiation weakly polarized
in a direction parallel to the emission surface, an orientation we
call {\it horizontal} polarization \cite{chandra:60}. At higher
energy, corresponding to 
the higher temperature of the inner disk, a greater fraction of the
emitted photons returns to the disk and is then scattered to the
observer. These scattered photons have a high degree of polarization
and are aligned parallel to the disk rotation axis ({\it vertical}),
as projected onto the image plane. At the transition point between
horizontal and vertical polarization, the relative contributions of
direct and reflected photons are nearly equal, and little net polarization is
observed. Since the effects of returning radiation are greatest for
photons coming from the innermost regions of the disk, the predicted
polarization signature is strongly dependent on the behavior of gas
near and inside the inner-most stable circular orbit (ISCO). Thus,
polarization observations could be used to measure the spin of the
black hole.

\section{Direct Radiation}
\begin{figure}
\centering
\scalebox{0.55}{\includegraphics*[52,450][355,720]{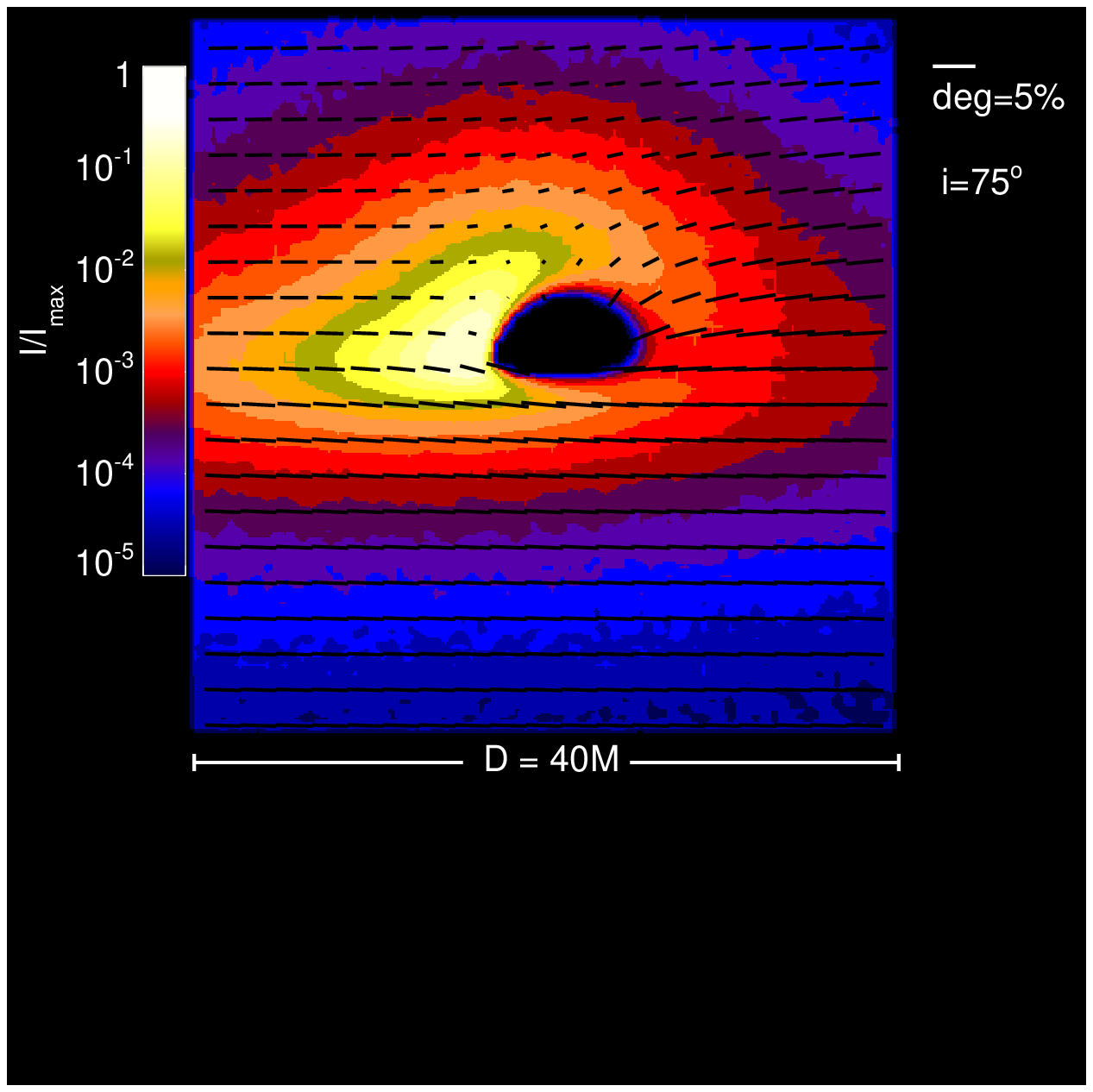}}
\scalebox{0.55}{\includegraphics*[115,450][410,720]{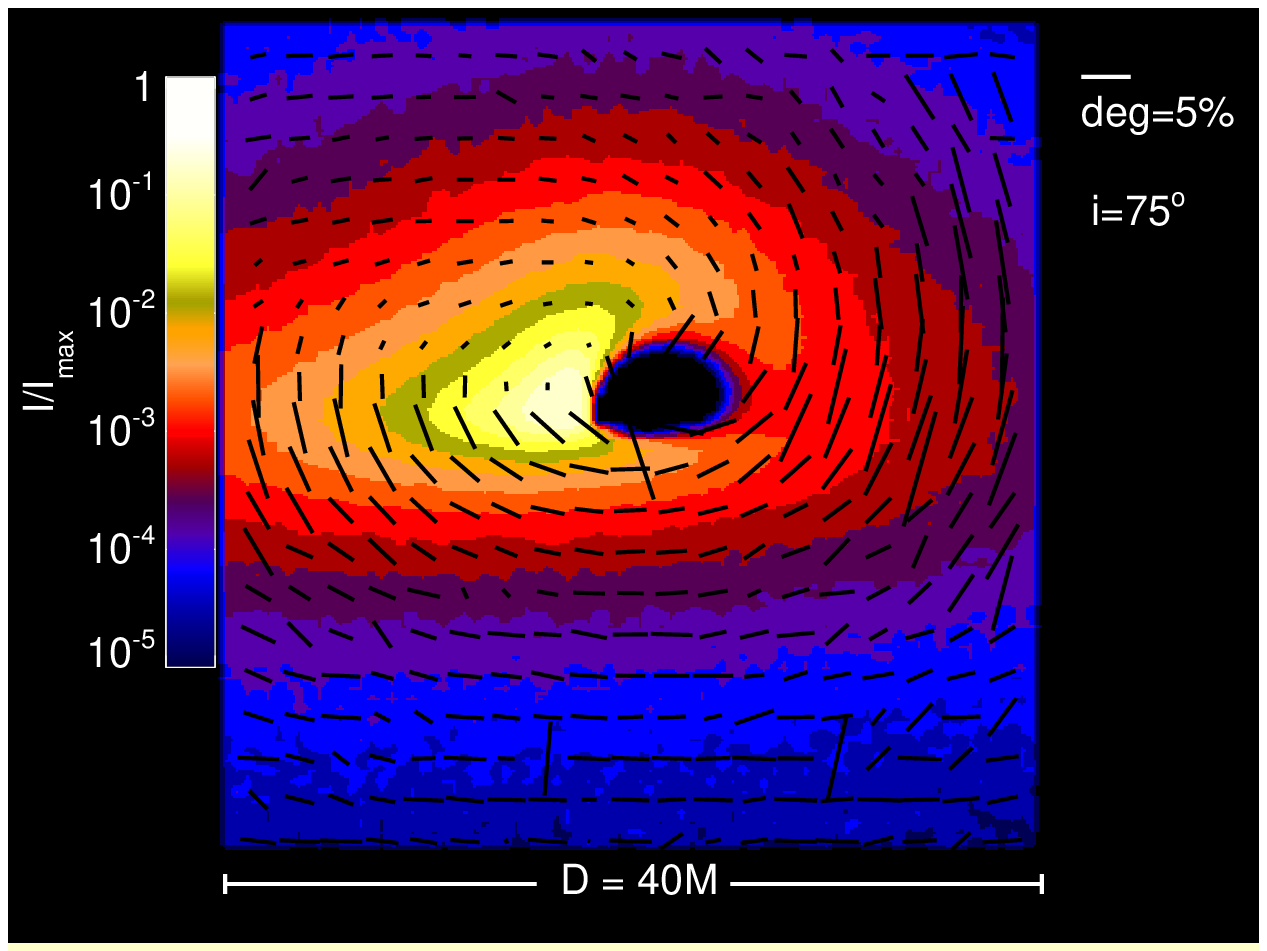}}
\caption{Ray-traced image of a thermal disk, including
  ({\it left}) only the direct, as well as ({\it right})
  both direct and returning radiation. The observer is located at an inclination of
  $75^\circ$ relative to the BH and disk rotation axis, with the gas
  on the left side of the disk moving towards the observer.
  The black hole has spin $a/M=0.9$, mass $M=10 M\odot$.}
\label{direct_image}
\end{figure}

In Figure \ref{direct_image}a, we show a simulated image of a
Novikov-Thorne \cite{novikov:73} accretion disk around a black hole
with spin parameter $a/M=0.9$ and luminosity $0.1 L_{\rm Edd}$,
corresponding to a disk whose X-ray spectrum peaks around 1 keV for a
BH mass of $10 M_\odot$. With the observer at an
inclination of $i=75^\circ$, significant relativistic
effects are clearly apparent. The increased intensity on the left side of
the disk is due to special relativistic beaming of the gas moving
towards the observer, and the general relativistic light bending makes
the far side of the disk appear warped and bent up towards the
observer. Superposed on top of the intensity map is the polarization
signature, represented by small black vectors whose lengths are
proportional to the degree of polarization observed from that local
patch of the disk.  Far from the
black hole, the polarization is essentially given by the classical
result of Chandrasekhar \cite{chandra:60} for a scattering-dominated atmosphere:
horizontal orientation with $\delta \approx 4\%$ when $i=75^\circ$;
nearer to the black hole, a variety of relativistic effects alter
the polarization.

The two most prominent relativistic effects are gravitational lensing
and special relativistic beaming, both lowering the net level of
polarization seen by the observer. Gravitational lensing causes the
far side of the disk to appear warped up towards the observer, and
thus have a smaller effective inclination.  Relativistic beaming causes
photons emitted normal to the disk plane in the fluid frame to travel
forward in the direction of the local orbital motion when seen by
a distant observer; the result is a smaller effective
inclination and thus degree of polarization.
Naturally, these relativistic effects are most
important close to the black hole, where the gas is also hottest and
the photons have the highest energies. All these effects are clearly visible in
Figure \ref{direct_image}a, which shows a smaller degree of
polarization where the beaming is greatest (yellow region of high
intensity in the left of the image) and the lensing is strongest (just
above the center of the image).

\section{Returning Radiation}

When returning radiation is included, although little
changes in terms of the total observed spectrum, the polarization picture
(Fig.~\ref{direct_image}b) changes significantly---in much of
the disk, the observed polarization rotates by $90^\circ$,
even though none of the model's physical parameters has
been changed at all!

This effect can be understood qualitatively in very simple
fashion (see also \cite{agol:00}). Since the reflected flux can be
roughly approximated by a single Thomson scatter, the polarization of
forward-scattered photons remains unchanged, while for
nearly-perpendicular scattering, the outgoing light 
can be close to $100\%$ polarized, in a direction normal to the
scattering plane.

For observers at high inclination angles, such as in
Figure~\ref{direct_image}b, returning radiation photons initially emitted
from the far side of the disk (top of the image) are reflected off the
near (bottom) side with a relatively small scattering angle,
maintaining a moderate horizontal polarization as in
Figure~\ref{direct_image}a.
On the other hand, photons emitted from the left side of the disk can be
bent back to the right side (or {\it vice versa}), and then scatter at
roughly $90^\circ$ to reach the observer, thereby aquiring a large vertical
polarization component. 
Although relatively small in total flux, this latter contribution
can dominate the polarization because it is so strongly polarized.

Because high-energy photons
from the hotter inner parts of the disk experience stronger
gravitational deflection, they are more likely to return to the
disk than the low-energy photons emitted at larger radii. We find
that, for each value of the spin parameter, there is a
characteristic ``transition radius,'' within which the returning
radiation dominates and produces net vertical polarization. Outside 
this point, the direct radiation dominates and
produces horizontal polarization. This transition
radius is in fact only weakly dependent on $a/M$, ranging from
$R_{\rm trans} \approx 7M$ for $a/M=0$ to $R_{\rm trans} \approx 5M$ for
$a/M=0.998$. The location and shape of the observed polarization swing 
can be used to infer the radial temperature profile, and thus the
spin, near the transition radius.

\begin{figure}
\centering
\includegraphics[width=0.48\textwidth]{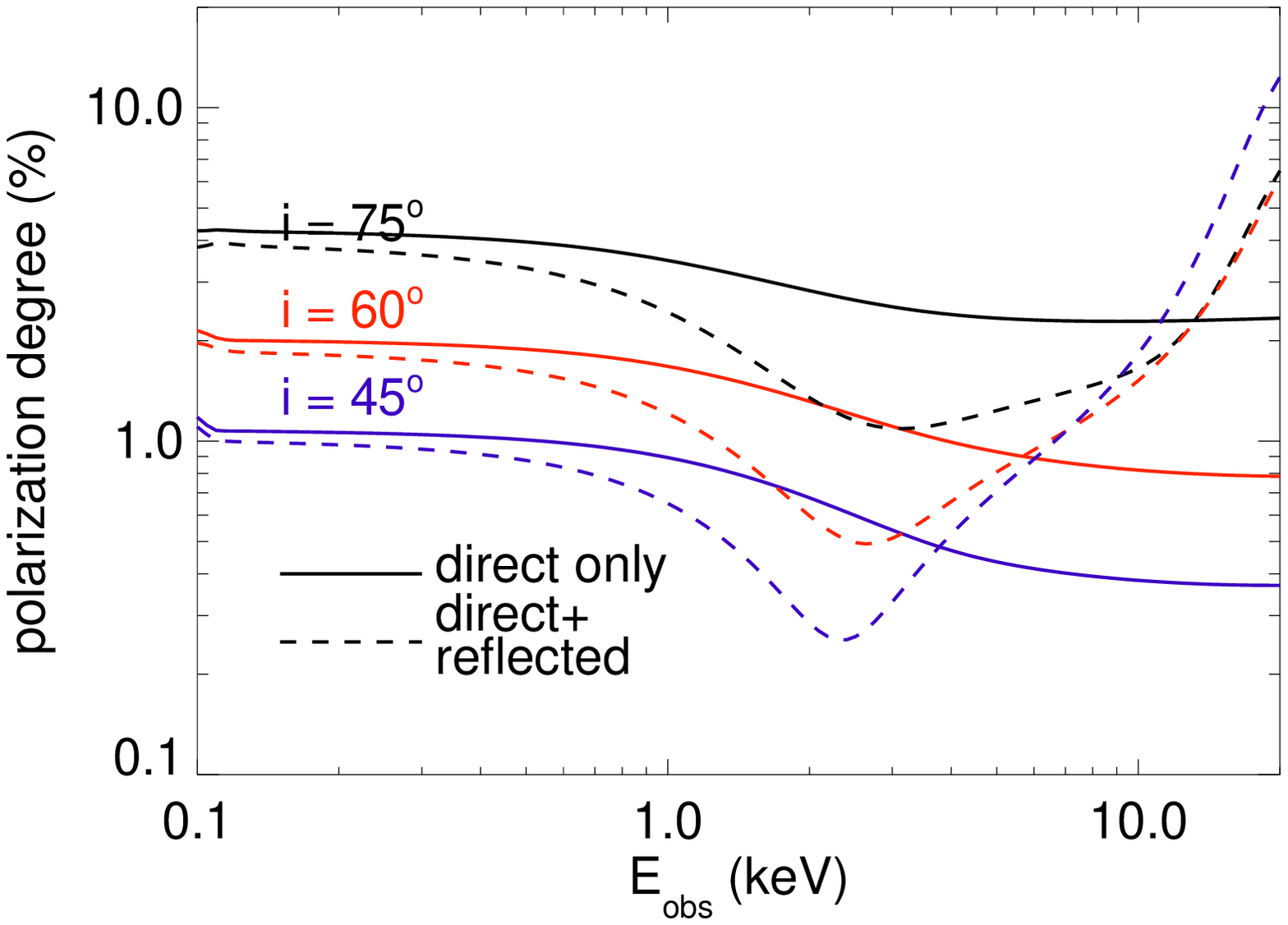}
\includegraphics[width=0.48\textwidth]{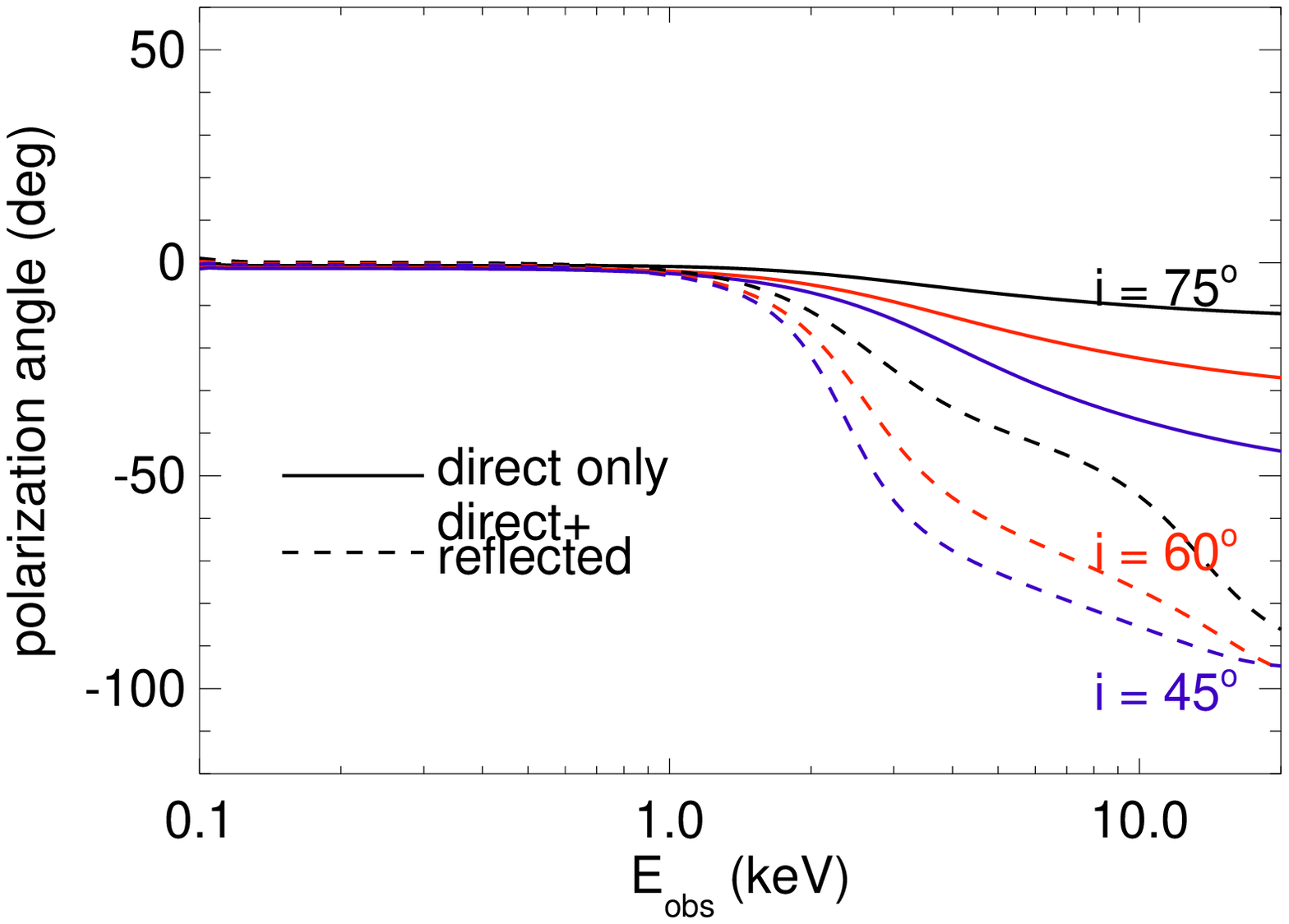}
\caption{Polarization degree ({\it left}) and angle ({\it right}) for
  a $10 M_\odot$ BH with spin $a/M=0.9$ and luminosity $L=0.1 L_{\rm
  Edd}$, for both the direct radiation case ({\it solid curves}) and
  also when including return radiation ({\it dashed curves}).}
\label{direct_return}
\end{figure}

\begin{figure}
\centering
\includegraphics[width=0.48\textwidth]{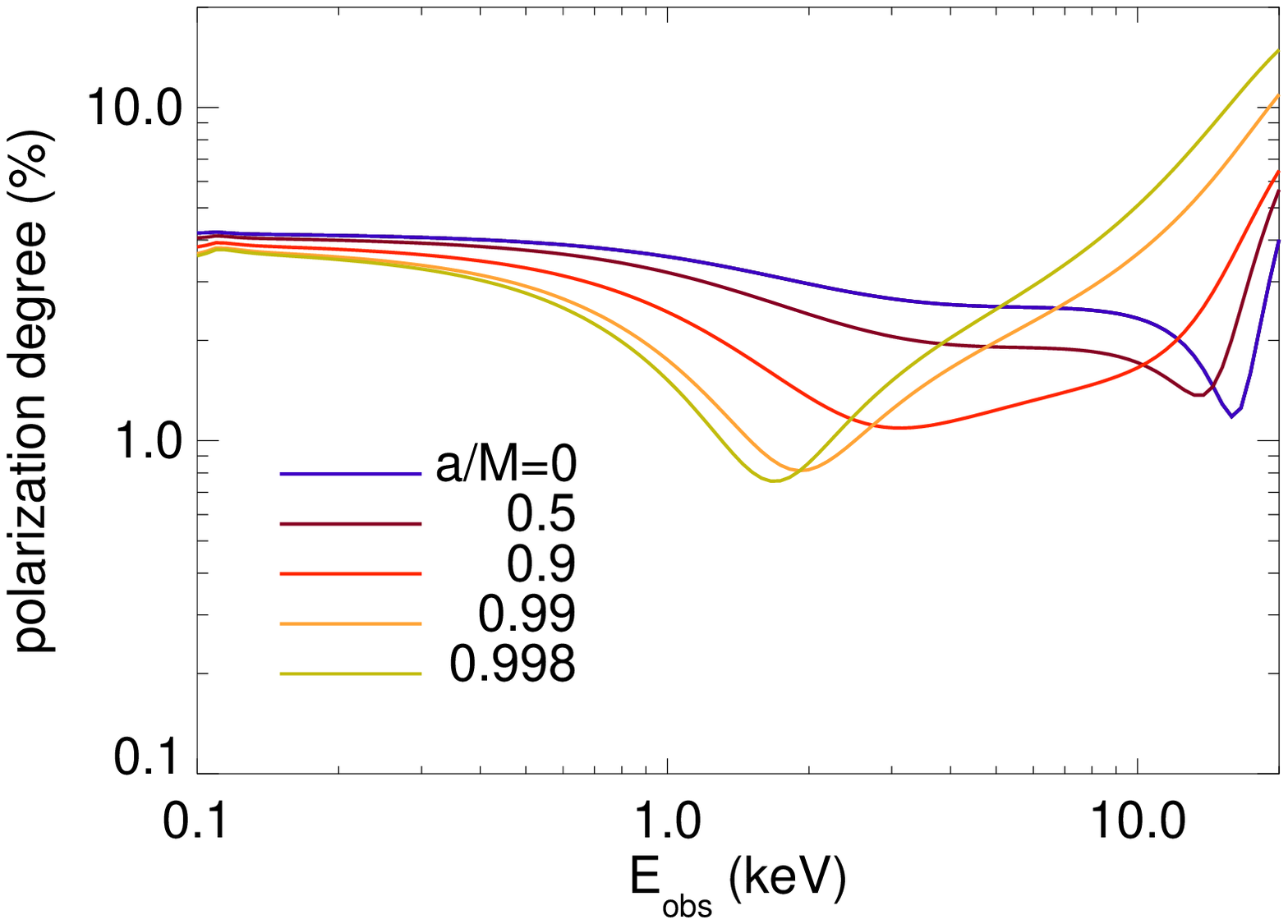}
\includegraphics[width=0.48\textwidth]{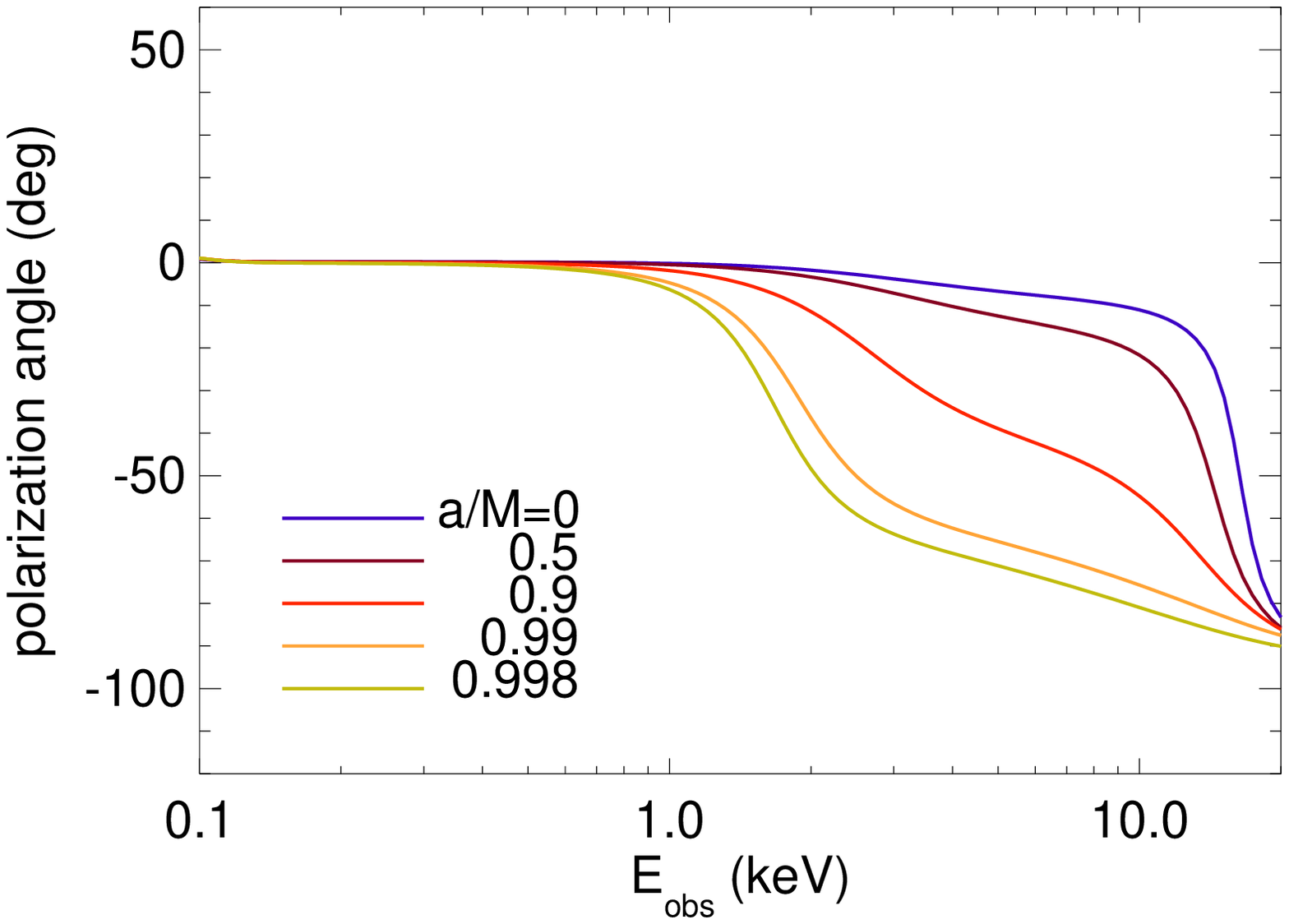}
\caption{Polarization degree ({\it left}) and angle ({\it right}) for
  a $10 M_\odot$ BH with luminosity $L=0.1 L_{\rm Edd}$, viewed from
  an inclination of $75^\circ$. The emissivity model is that of
  Novikov-Thorne for a range of BH spins.}
\label{spin2}
\end{figure}

In Figure \ref{direct_return}, we compare the observed polarization as
a function of energy for inclinations $i=45, 60, 75^\circ$, considering
only direct radiation (solid curves) and also including return
radiation (dashed curves). Clearly, when including the reflected
photons, the polarization swings from horizontal at low energies to
vertical above the thermal peak. 
In Figure \ref{spin2} we show the polarization degree and angle for a
range of BH spin parameters, in all cases including return
radiation. As the spin increases, the ISCO moves
closer to the horizon and a greater portion of the flux is emitted
inside the transition radius, resulting in a higher fraction of return
radiation. Thus more of the flux reaching the observer is vertically
polarized, and the energy of transition moves closer to the thermal
peak. 

\section{Discussion}
As described in \cite{schnittman:09}, we have developed a new
ray-tracing code to calculate the X-ray polarization from accreting
black holes. In this paper we focus on polarization signatures of the
thermal state in stellar-mass black holes. The emitted radiation has a
diluted thermal spectrum and is weakly polarized parallel to the disk
surface. The integrated polarization spectrum seen by a distant
observer contains
distinct energy-dependent features, as the high-energy photons from
the inner disk are modified by relativistic effects such as Doppler
boosting and gravitational lensing near the BH.

For radiation originating very close to the BH, the most important
relativistic effect is the strong gravitational lensing that causes
the photons to get bent back onto the disk and scatter towards the
observer.  This scattering can induce very high levels of polarization,
and leads to a distinct
transition in the polarization angle from horizontal at low energies
to vertical above the thermal peak. Observing such a swing in
the polarization angle would give the most direct evidence to date for
the extreme relativistic light bending predicted around black
holes. 
By measuring the location and shape of the polarization
transition, we will be able to constrain the temperature profile of the
inner disk.  Assuming a Novikov-Thorne disk truncated at
the ISCO, the polarization signal
gives a direct measurement of the BH spin.

\end{document}